# Remote optical addressing of single nano-objects

M. Brun, A. Drezet, H. Mariette, J.C. Woehl,  S. Huant [*]

*Laboratoire de Spectrométrie Physique, CNRS UMR5588, Université Joseph Fourier*
*Grenoble - BP 87, 38402 Saint Martin d'Hères cedex, France*

* E-mail: serge.huant@ujf-grenoble.fr

**PACS**

07.79.Fc -Near-field scanning optical microscopes.

78.67.-n - Optical properties of nanoscale materials and structures.

73.20.Mf - Collective excitations (surface/interface states) including excitons, polarons, plasmons and other charge-density excitations

We present a scheme for remotely addressing single nano-objects by means of near-field optical microscopy that makes only use of one of the most fundamental properties of electromagnetic radiation: its polarization. A medium containing optically active nano-objects is covered with a thin metallic film presenting sub-wavelength holes. When the optical tip is positioned some distance away from a hole, surface plasmons in the metal coating are generated which, by turning the polarization plane of the excitation light, transfer the excitation  towards a chosen hole and induce emission from the underlying nano-objects. The method, easily applicable to other systems, is demonstrated for single quantum dots (QDs) at low temperature. It may become a valuable tool for future optical applications in the nanoworld.

Ultimate control of light requires the combined ability of confining photons to extremely small dimensions, *i.e.* much smaller than their wavelength, and manipulating them in a well-controlled state to address at will optically active single nanometer-scaled objects, such as single molecules, semiconductor QDs or nanocrystals. Routes in this direction have been opened up by Near-field Scanning Optical Microscopy (NSOM) [1] : for instance, early single molecule detection [2] with NSOM using conventional optical tips has recently evolved to NSOM imaging with a single molecule serving as a point-like light source [3]. Another example is the quantum corral [4], an elegant realization of Scanning-Tunneling Microscopy, which has been extended recently to the optics world by NSOM [5]. Here, we present a simple optical method aimed at remotely addressing single nano-objects by means of NSOM. Our method does not need sub-wavelength waveguides as a prerequisite [6]. In essence, it makes only use of one of the most fundamental properties of an electromagnetic field with respect to a scalar field: its polarization state, which is controlled at the apex of a NSOM tip [7]. We demonstrate our approach on semiconductor QDs at low temperature; however, there is no restriction whatsoever to apply it to a large variety of optically active nano-objects under various conditions, in particular under ambient conditions.

The principle of the method is sketched in fig. 1. A thin opaque metal film, hollowed by sub-wavelength apertures, is deposited onto a medium containing optically active nano-objects. A NSOM tip, *i.e.* a tapered and metal coated optical fiber with an optical aperture of typically 100 nm at the tip apex [8], is coupled to a laser source and is approached to the surface with 3D control of its position with an accuracy of 10 nm or better, as is commonly done in nowadays NSOM [9]. This allows light to be transmitted through the tip directly to a selected nanohole or to a well-controlled position on the metal film. The latter possibility is particularly promising since orienting the light polarization at the apex of the optical tip towards a particular hole allows to selectively launch surface plasmons – 2D collective excitations of the electron plasma in the metal film – towards that particular hole. Subsequent plasmon scattering at the hole boundaries or/and on the back side of the film allows to recover the plasmon energy under the form of light with the same wavelength as the incident light [10]. We demonstrate below that this selective launching and scattering of 2D plasmons allows for a remote, surface-plasmon-mediated addressing of the underlying optically active medium.

In our demonstration, the active medium is made of self-assembled CdTe semiconductor QDs grown by Atomic-Layer Epitaxy [11]. The growth sequence includes a ZnTe buffer layer, a 2.1 nm thick CdTe layer, and a protecting 30 nm ZnTe cap layer. In this CdTe/ZnTe system, carrier confinement to zero dimension is ensured by Cd-rich islands (typical density of 5 x $10^{10}$ cm$^{-2}$, average height and width of 7 nm and 15 nm, respectively) that form due to elastic relaxation of the strained CdTe layer. A hollowed metal film is deposited onto this semiconductor surface according to the following "lift-off" procedure. We first disperse commercially available latex beads with a diameter of 240 nm on the surface, then evaporate a 70 nm thick aluminum layer (sufficient to ensure opacity to visible light) by vacuum evaporation, and finally dissolve the latex nanospheres in an organic solvent. Optical information is subsequently gained from this QD medium by far field collection of the low-temperature (4.2 K) luminescence emitted by single CdTe QDs [12] located underneath the selected nanohole. The low-temperature NSOM microscope used in this study has been described elsewhere [9].

Figure 2a) shows a 2 µm x 2 µm reflection image exhibiting two dark spots on its upper and left borders due to two nanoholes located at these positions. These two apertures are labelled H1 and H2, respectively. As shown in fig. 2b), the H1 nanohole luminescence spectrum, *i.e.* the luminescence spectrum recorded with the optical tip exciting the upper hole position (H1), consists of a set of very sharp lines that are typical for the atomic-like shell structure [12-14] of the underlying dots. In the most general case, these sharp peaks are due to radiative recombinations of various multi-exciton states, either electrically neutral [13] or negatively charged [14], that can take place in several close-lying QDs. This luminescence information is unique for a given set of dots which vary in size, composition, and local density [11]. As a consequence, it forms a spectral fingerprint of a particular nanohole in the metal layer that is unique for this particular hole.

In order to understand the effect of rotating the light polarization on the QD luminescence, we first need to find a reliable method to determine *in situ* this polarization. This is achieved by analyzing the high-contrast reflection image in fig. 3 where 3 nanoholes appear as three-lobe structures made of two bright spots separated by a marked, slightly elongated, dark spot. This identifies the light polarization as being perpendicular to the axis joining the two bright spots, as revealed by the simulation in fig. 3c), where the electromagnetic field created by an optical tip has been viewed as emanating from two crossed radiating dipoles located in the tip aperture plane, namely an electric dipole $\mathbf{P}_{tip}$ and a magnetic dipole $\mathbf{M}_{tip}$, with the electric dipole aligned with the incident polarization [7,15]. In the case of the optical tip facing the aluminum coating which is considered as a perfect metal, the reflected image dipoles are such that $\mathbf{P}_{image} = -\mathbf{P}_{tip}$ and $\mathbf{M}_{image} = \mathbf{M}_{tip}$. As a result, the total field acting on a hole situated in the vicinity of the tip is a magneto-static dipolar field only, to which the hole reacts like an effective magnetic dipole $\mathbf{M}_{hole}$ [7,16]. The recorded far-field intensity is directly proportional to $\left|2\mathbf{M}_{\text{tip}}+\mathbf{M}_{\text{hole}}\right|^2$. This yields the simulated image of fig. 3c), which is in qualitative agreement with the experiment. In the simulation, the hole itself has been viewed as being perfectly transparent. Rotating the excitation polarization in the optical fiber by means of a paddle-polarization-controller arranged in the optical path rotates the 3-lobe structures of fig.3 accordingly, thereby providing us with the desired *in-situ* determination of light polarization.

To investigate the light polarization effect on the luminescence spectra, two intense peaks are isolated from the reference spectrum of fig. 2b) and their intensity is analyzed as a function of the excitation polarization (the reference polarization state, named "0°-polarization" in the following, is marked by an arrow on the H2 nanohole, fig. 2a)). As seen in fig. 2c), the polarization rotation has almost no effect on the luminescence intensity when the tip is located directly over the H1 nanohole (point 1). This is because in the absence of external fields, the CdTe QDs are unpolarized in the sense that they provide isotropic emission. However, the situation changes in a dramatic way as soon as the tip is positioned further away from the nanohole as shown on fig. 2d) and 2e) for tip positions 2 and 5, respectively. Here, the experiment clearly reveals a preferential polarization state that enhances the luminescence intensity.

Our results are summarized on fig. 2a) where an arrow indicates the particular direction that magnifies the corresponding nanohole response for each of the four studied points around nanohole H1. A clear trend is revealed in fig. 2a): the hole signature is enhanced for electric field orientations pointing essentially towards the hole, irrespective of the launching point. Remaining small offsets can arise from plasmon scattering due to metal roughness. This demonstrates that the excitation necessary for stimulating emission from the nanohole is transported to the latter on the surface of the metal film which is opaque to incident light.

A demanding test of the surface plasmon scenario already anticipated comes from the belavior of the luminescence intensity. Within this scenario, the collected luminescence intensity should be proportional to the surface-plasmon intensity $I_{SP} \propto \left|\Psi(\rho,\phi,z)\right|^2$ at the hole position, which depends both on the tip-hole distance $\rho$ and on the polarization direction $\phi$. $I_{SP}$ can be computed writing $\Psi(\rho,\phi,z)=\cos(m\phi)e^{\pm kz}H_m\left(\chi\rho\right)$ for the typical 2D plasmon-mode function in a cylindrical geometry where the complex wavevector components $k$ (axial) and $\chi$ (radial) are linked by the frequency-dependent medium permittivity $\varepsilon(\omega)$ in a general dispersion relation of the type $\left(\omega\!/\!c\right)^2\varepsilon(\omega)=-k^2+\chi^2$. In the above expression for $\Psi$, an obvious time dependence of the form $e^{i\omega t}$ has been omitted, $m$ is an integer with $m = 1$ for the ground plasmon mode and $H_m(x)$ is the $m^{th}$ Hankel function [17]. Writing the boundary conditions for the plasmon at the aluminum-air interface further implies $\varepsilon(\omega) = -\frac{k}{k_1}$, where $k$ is the z-wavevector of the plasmon in aluminum and $k_1$ that in the air. The present simple analysis gives a (far-field) attenuation length of surface plasmons in aluminum of $L_{SP} = 4.2\mu m$ at an optical wavelength of 515 nm for which $\varepsilon = -38.7938 + 10.3795i$ [18]. It is interesting to note that the far-field radial asymptotic behavior $I_{SP} \propto \frac{e^{-\rho/L_{SP}}}{\rho}$ [19] is recovered because at large distance, $H_1(\chi\rho) \to \frac{e^{-\chi\rho}}{\sqrt{\rho}}$, while the angular dependence of $I_{sp}$ directly reflects the polarization of the electric near-field at the tip apex or, in an equivalent manner, the charge distribution at the tip apex [20]. It takes the simple functional form of $\cos^2(\phi)$.

Figure 4 compares the angular and radial dependences of the luminescence intensity with the expected behavior for surface plasmons (here, we only consider the low-energy component of the doublet discussed on fig. 2, the high-energy component and other peaks behave in the same way). The agreement is excellent for the actual material parameters in aluminum, which gives strong support to the surface plasmon interpretation. In "plasmon metals", such as silver, or for excitation light close to the plasmon resonance, the plasmon attenuation length is much larger [19] which would allow for object addressing from more remote tip positions. However, the lateral propagation of surface plasmons, which manifests itself in the $\phi$ dependence of $I_{sp}$, may in turn cause limitations in the addressing accuracy. Therefore, aluminum offers a good compromise for the purpose of a first demonstration.

We have introduced a simple optical way of remotely addressing single nano-objects. This method makes use of the control of light polarization at the apex of an optical tip to direct at will the excitation into sub-wavelength apertures hollowed into a thin metal coating deposited onto the active medium. The excitation is mediated by surface plasmons. Therefore, it can be launched from any point to any other point within the attenuation length of surface plasmons. As a consequence, the objects to address may be dispersed almost freely on the surface, or even below the metal surface like in the present experiment, providing that some indivuals remain accessible through a nanohole. Since most of the active objects placed directly on the metal surface are expected to quench due to the strong interaction of their transition dipole with the image dipole in the metal, the method should be only sensitive to those objects located in or just underneath a nanohole. Other promising features of our approach include: i) the nanohole diameter fits with the initial latex-bead diameter and can be tuned accordingly over a wide range; ii) the average distance between neighboring holes can be controlled freely by adequatly choosing the bead concentration; iii) the initial nanospheres can be dispersed onto nanostructured surfaces which, as can be anticipated, may help in arranging the final nanoholes in a well-defined manner. This could lead to extraordinary cooperative effects [21]. Furthermore, various optically active nano-objects can be addressed, such as single molecules [2,3,22] or nanocrystals [23], not only QDs at low temperature. Therefore, we believe that our method opens a way to a large variety of optical experiments in the nanoworld, with possible applications for the optical addressing of quantum bits and experiments with single and entangled photons [24].

We thank stimulating discussions with R. Romestain, M. Stark and J.C. Vial. We gratefully acknowledge support from the Institut de Physique de la Matière Condensée - Grenoble, the Région Rhône-Alpes, the CNRS, and the Volkswagen Foundation.

# References

[1] Pohl D. W. and Lanz W. D. M., *Appl. Phys. Lett.*, **44**, (1984) 651.

[2] Betzig E. and Chichester R., *Science,* **262**, (1993) 1422 .

[3] Michaelis J., Hettich C., Mlynek J. and Sandoghdar V., *Nature,* **405**, (2000) 325.

[4] Crommie M. F., Lutz C. P. and Eigler D. M., *Science,* **262**, (1993) 218.

[5] Chicanne C. *et al.*, *Phys. Rev. Lett.,* **88**, (2002) 097402.

[6] Colas des Francs G., Girard Ch., Weeber J. C. and Dereux A., *J. Microscopy,* **202**, (2001) 307.

[7] Obermüller C. and Karrai K., *Appl. Phys. Lett.,* **67**, (1995) 3408.

[8] Betzig E., Trautman J. K., Harris T. D., Wliner J. S. and Kostelak R. L., *Science,* **251**, (1991) 1468.

[9] Brun M. *et al.*, *J. Microscopy,* **202**, (2001) 202.

[10] Sönnichsen C. *et al., Appl. Phys. Lett.*, **76**, (2000) 140.

[11] Marsal L. *et al.*, *J. Appl. Phys.*, **91**, (2002) 4936.

[12] Brun M. *et al., Solid State Commun.,* **121**, (2002) 407.

[13] Bayer M., Stern O., Hawrylak P., Fafard S. and Forchel A*., Nature,* **405**, (2000) 923.

[14] Warburton R. J.  *et al.*, *Nature,* **405**, (2000) 926.

[15] Drezet A., Woehl J. C. and Huant S., *Europhys. Lett.,* **54**, (2001)736.

[16] Bouwkamp C. J., *Philips Res. Rep.,* **5**, (1950) 321.

[17] Gradshteyn I. S. and Ryzhik I. M., *Table of Integrals, Series, and Products, 6$^{th}$ edition* (Academic Press, San Diego & London) 2000.

[18] Novotny L., thesis, Swiss Federal Institute of Technology, Zurich (1996).

[19] Hecht B., Bielefeldt H., Novotny L., Inouye Y. and Pohl D.W., *Phys. Rev. Lett.* **77**, (1996) 1889.

[20] Drezet A., Huant S. and Woehl J.C., *to be published.*

[21] Ebbesen T. W., Lezec H. J., Ghaemi H. F., Thio T. and Wolff  P. A., *Nature*, **391**, (1998) 667.

[22] Moerner W.E. and Orrit M., *Science*, **283**, (1999) 1670.

[23] Naber A. *et al.*, *Phys. Rev. Lett.,* **89**, (2002) 210801.

[24] Altewischer E., van Exter M. P. and Woerdman J. P., *Nature*, **418**, (2002) 304.

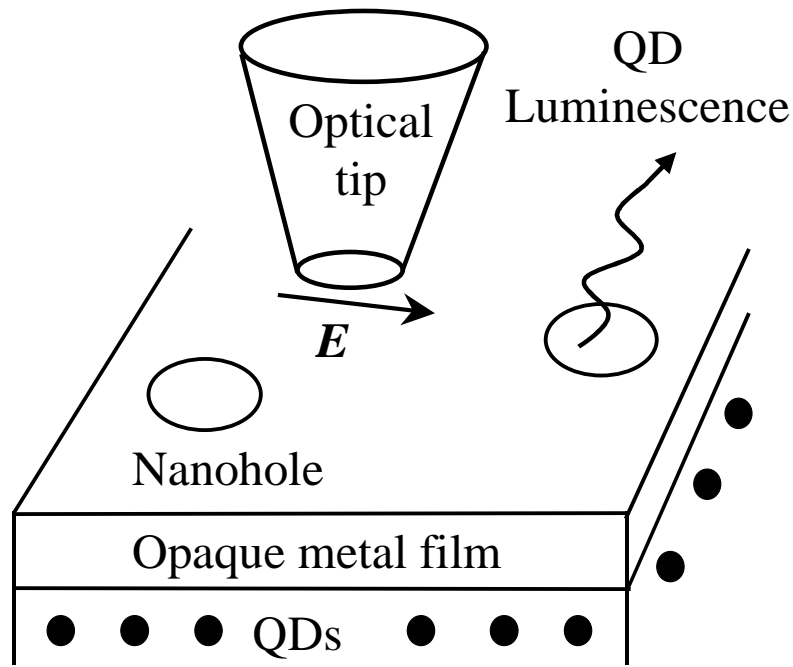

Fig.1- Principle of the remote addressing method. The optical excitation **E** emerging from the optical tip is polarized towards a selected nanohole in an opaque metal film to which it is transfered by surface plasmons. The hole signature is obtained from the luminescence light emitted from underlying QDs, represented here as closed circles.

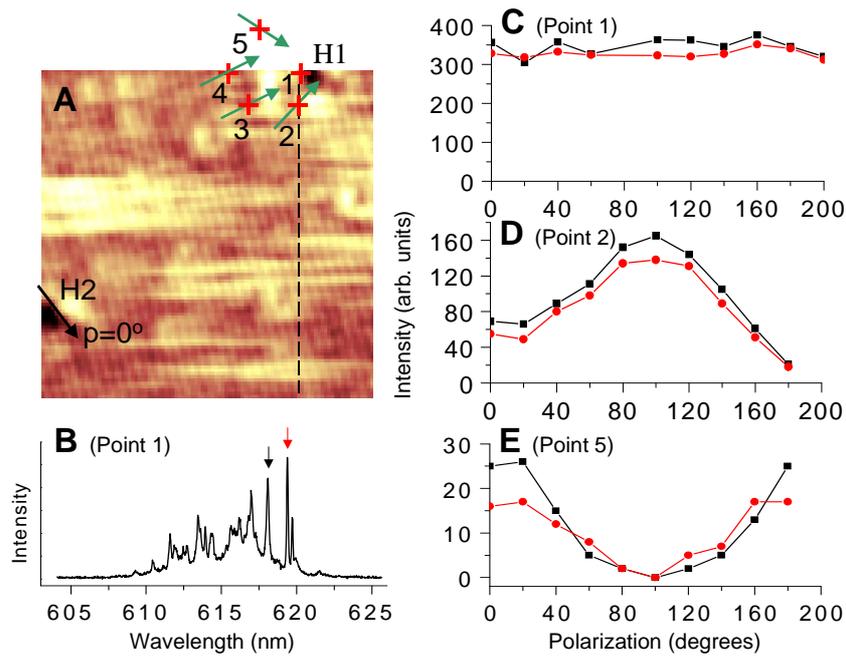

Fig. 2- Demonstration of the addressing method. a) A 2 µm x 2 µm reflection image of the hollowed mask with two nanoholes on the upper and left borders (dark spots). The 0° reference polarization state is marked by an arrow on the H2 nanohole. The four green arrows pointing towards nanohole H1 give the polarization directions for each addressing point 2 to 5 with the highest luminescence response collected from nanohole H1. b) A luminescence reference spectrum (T= 4.2 K) with the optical tip located over nanohole H1 (point 1). The light emitted at 514.5 nm by an Argon ion laser is coupled into the tip and excites the sample in the near-field. The emission is collected in the far-field above the sample, dispersed through a monochromator and analyzed by a nitrogen-cooled CCD camera. c) to e) Luminescence intensity of the two sharp peaks of fig. 2b) marked by a black and a red arrow as a function of the polarization direction for points 1, 2 and 5, respectively. All three plots use the same intensity scale. The power injected into the tip is 400 µW (transmission of 8 x 10$^{-4}$) and the integration time is 30 s per spectrum.

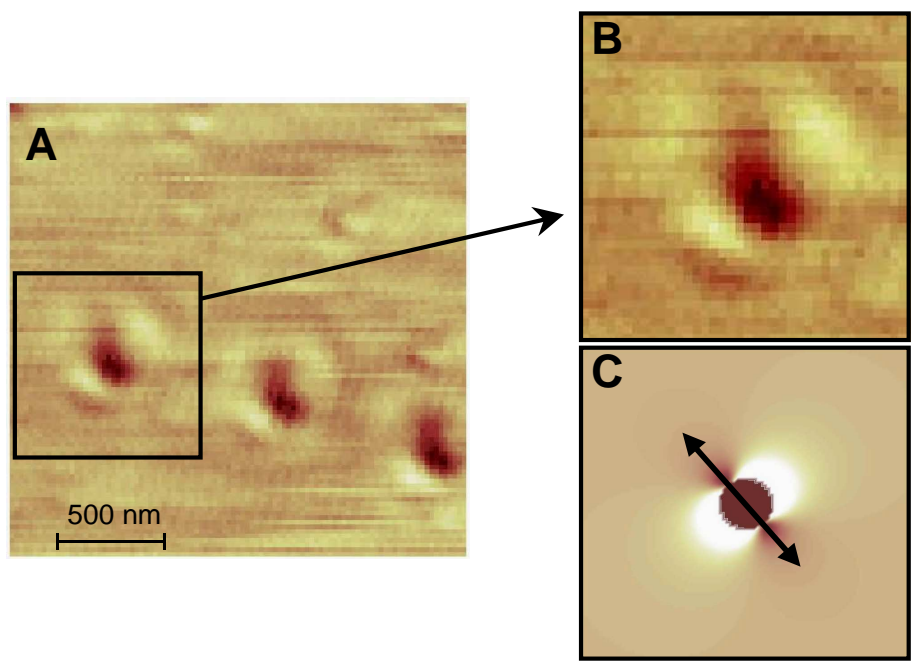

Fig. 3- A method for determining *in situ* the light polarization. a) A 2 µm x 2 µm reflection image of the hollowed mask with three nanoholes seen as slightly elongated dark spots accompanied by two brighter spots. b) A zoom on the left hole of fig 3a) compared with, c) a simulation as described in the text. The double arrow in fig. 3c) indicates the polarization direction in the simulation.

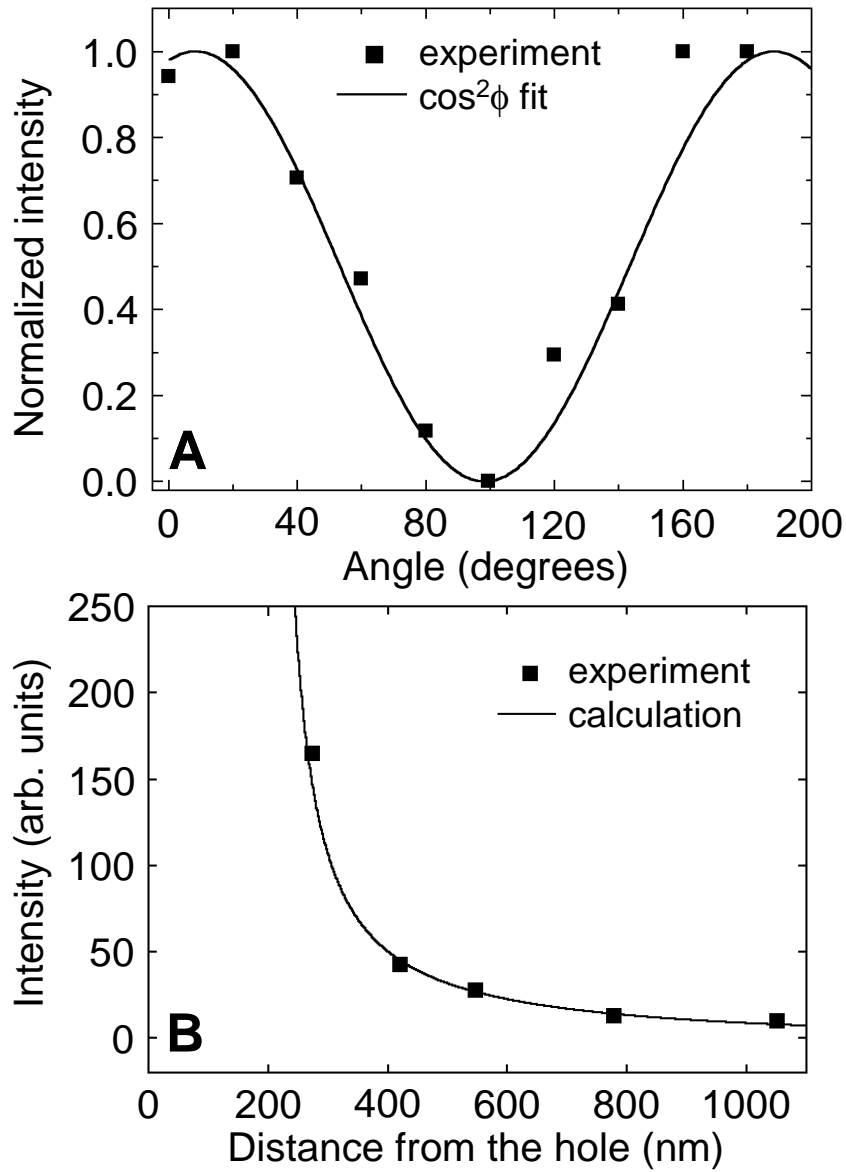

Fig. 4- Luminescence intensity *versus* surface plasmon intensity at the nanohole position. a) Angular dependence of the luminescence intensity. The experimental data correspond to the low-energy component of the intense doublet of fig. 2b) with the optical tip located on point 5 of fig. 2a). The background signal detected by the CCD camera has been removed and the measured intensity has been normalized to its maximum. b) Radial dependence of the luminescence intensity. The experimental data are for the same sharp peak as in fig. 4a): they are collected along the dashed line of fig. 2a) with the electric excitation field pointing towards the nanohole H1. The power injected into the fiber tip and the integration time are the same as in fig. 2. The calculated curve (see text) has been shifted to the right by an offset value of 200 nm, which accounts for the finite hole diameter (240 nm) and tip aperture (≈ 100 nm).